# RR-Lyrae-type pulsations from a 0.26-solar-mass star in a binary system

G. Pietrzyński[1,2], I.B. Thompson[3] ,W. Gieren[1], D. Graczyk[1], K. Stępień[2] , G. Bono[4,5], P. G. Prada Moroni[6,7] , B. Pilecki[1,2], A. Udalski[2], I. Soszyński[2], G. Preston[3] , N. Nardetto[8], A. McWilliam[3], I. Roederer[3], M.. Górski[1,2] , P. Konorski[1,2] J. Storm[9]

1. *Universidad de Concepción, Departamento de Astronomìa, Casilla 160-C, Concepciòn, Chile*
2. *Obserwatorium Astronomiczne Uniwersytetu Warszawskiego, Aleje Ujazdowskie 4, 00-478 Warszawa, Poland*
3. *Carnegie Observatories, 813 Santa Barbara Street, Pasadena, CA 911101-1292, USA*
4. *Dipartimento di Fisica Universita' di Roma Tor Vergata, via della Ricerca Scientifica 1, 00133 Rome, Italy*
5. *Dipartimento di Fisica Universita' di Pisa, Largo B. Pontecorvo 2, 56127 Pisa, Italy*
6. *INFN, Sez. Pisa, via E. Fermi 2, 56127 Pisa, Italy*
7. *INAF-Osservatorio Astronomico di Roma, Via Frascati 33, 00040 Monte Porzio Catone, Italy*
8. *Laboratoire Lagrange, UMR7293, UNS/CNRS/OCA, 06300 Nice, France*
9. *Leibniz Institute for Astrophysics, An der Sternwarte 16, 14482, Postdam, Germany*

**RR Lyrae pulsating stars have been extensively used as tracers of old stellar populations for the purpose of determining the ages of galaxies, and as tools to measure distances to nearby galaxies[1,2,3]. There was accordingly considerable**

**interest when the RR Lyr star OGLE-BLG-RRLYR-02792 was found to be a member in an eclipsing binary system[4], as the mass of the pulsator (hitherto constrained only by models) could be unambiguously determined. Here we report that RRLYR-02792 has a mass of 0.26 $M_\odot$ and therefore cannot be a classical RR Lyrae star. Through models we find that its properties are best explained by the evolution of a close binary system that started with 1.4 $M_\odot$ and 0.8 $M_\odot$ stars orbiting each other with an initial period of 2.9 days. Mass exchange over 5.4 Gyr produced the observed system, which is now in a very short-lived phase where the physical properties of the pulsator happen to place it in the same instability strip of the H-R diagram occupied by RR Lyrae stars. We estimate that samples of RR Lyr stars may contain a 0.2 percent contamination with systems similar to this one, implying that distances measured with RR Lyrae stars should not be significantly affected by these binary interlopers.**

Using high-resolution spectra obtained with the MIKE spectrograph at the 6.5 m Magellan Clay telescope at the Las Campanas Observatory in Chile, and the UVES spectrograph attached to the 8.2 m VLT telescope of the European Southern Observatory on Paranal (program 287.D-5022(A)), we confirmed that OGLE-BLG-RRLYR-02792 (hereafter: RRLYR-02792) is a true physical, well detached, double-lined eclipsing binary system very well suited for deriving the masses of its two components with a very high accuracy.

Analysis of the spectroscopic and photometric observations (see Figures 1 and 2) results in the determination of the astrophysical parameters of our system presented in Supplementary Table 1. Realistic errors of the derived system parameters were determined using Monte Carlo simulations. The resulting masses of the components turned out to be very unexpected. The dynamical mass of the RR Lyrae component, 0.261 ± 0.015 $M_\odot$, is much smaller than the mass required for helium ignition, and therefore completely at odds with the predictions of all theoretical models of RR Lyrae stars[5,6,7] Moreover, if the pulsating component of RRLYR-02792 were indeed a classical RR Lyrae star as suggested by its light curve and pulsation period, the nature of the more massive, cooler (at $T_1$ = 6000 K, $T_2$ = 0.68 × $T_1$) and (by some 2 magnitudes in V band) fainter secondary component would be extremely mysterious. Assuming a typical temperature for the RR Lyrae star of 6000 K, the temperature of the static secondary component would be only about 4100 K, much too cool for a giant star with $M_2$ = 1.67 $M_\odot$ (whose temperature is expected to be close to 5000 K).

A clue comes from the relatively short orbital period of 15.24 days which suggests that mass exchange between the two components should have occurred during the evolution of this system. Inspired by this possibility we calculated a series of models for Algol-type binary systems[8,9] and found that a system which initially contained two stars with $M_1$ = 1.4 $M_\odot$ and $M_2$ = 0.8 $M_\odot$ orbiting each other with an initial period of 2.9 days would, after 5.4 Gyrs of evolution, have exchanged mass between the components as classical Algols do, and today would form a system very similar to RRLYR-02792 (e.g. with $M_1$ = 0.268 $M_\odot$ and $M_2$ = 1.665 $M_\odot$ , and P=15.9 days).

We therefore conclude that the primary component of our observed system is not a classical RR Lyrae star with its well-known internal structure, but rather a star which possesses a partially degenerate helium core and a small hydrogen-rich envelope (shell burning) which has lost most of its envelope during the previous red giant branch phase to the secondary star due to mass exchange in the binary system, and which is now evolving towards the hot subdwarf region on the H-R diagram (see Figure 3 and Supplementary Figure 2). The pulsational light curve of such a star very closely resembles that of a classical RR Lyrae star. However, variable stars produced this way are expected to cross the classical pulsational instability region on the H-R diagram about a hundred times faster than the RR Lyrae stars. Because the star is moving rapidly at a constant luminosity across the instability strip toward higher temperatures, its radius should become smaller and therefore its pulsation period should steadily decrease. Indeed, using our photometric data we have measured a period decrease of the pulsating component of (8.4 ± 2.6) x $10^{-6}$ days/year, which is on average more than two orders of magnitude larger than the period change shown by canonical RR Lyrae stars[10], and therefore strongly supports our interpretation. Moreover, we have detected hydrogen lines in the spectrum of the binary associated with the pulsating primary component (see Supplementary Figure 1) which confirms that the star possesses a hydrogen-rich envelope. In such a scenario the secondary component is a typical red giant star currently evolving up the red giant branch increasing its size and luminosity. During its future evolution, our system will turn into a binary system composed of two white dwarfs sharing a common envelope.

We have captured the RRLYR-02792 binary system in a very special and short-lived phase of its evolution which constitutes just a small fraction ($10^{-4}$) of its current age. The system provides a number of strong observational constraints which have enabled us to unambiguously and in detail track its past evolution, and as a result discover a new evolutionary channel of producing binary evolution pulsating stars - new habitants of the pulsational instability strip on the H-R diagram which mimic classical RR Lyrae variables, but have a completely different origin. These low mass pulsating stars could in principle increase the observed spread in luminosity of the RR Lyrae stars, hence affecting distance measurements based on them. We roughly calculated that among 1000 RR Lyrae stars one should expect just 2 such stars, so in practice they should not affect distance determinations to galaxies if these are made with relatively large samples of RR Lyrae stars.

## REFERENCES

1. Mateo, M. Dwarf Galaxies of the Local Group. Annu. Rev. Astron. Astrophys. **36**, 435-506 (1998).

2. Bono, G. & Cignoni, M. Variable Stars as Standard Candles and Stellar Tracers. Proceedings of the Gaia Symposium "The Three-Dimensional Universe with Gaia" (ESA SP-576). Held at the Observatoire de Paris-Meudon, 4-7 October 2004. Editors: C. Turon, K.S. O'Flaherty, M.A.C. Perryman

3. Szewczyk, O., Pietrzynski, G., Gieren, W., Storm, J., Walker, A., Rizzi, L., Kinemuchi, K., Bresolin, F., Kudritzki, R.-P. & Dall'Ora M. The Araucaria Project. The Distance of the Large Magellanic Cloud from Near-Infrared Photometry of RR Lyrae Variables. Astron. J. **136**, 272-279 (2008).

4. Soszyński, I., Dziembowski, W., Udalski, A., Poleski, R., Szymański, M.K., Kubiak. M., Pietrzyński, G., Wyrzykowski, Ł., Ulaczyk., K., Kozłowski, S. & Pietrukowicz, P. The Optical Gravitational Lensing Experiment. The OGLE-III catalog of variable stars. XI. RR Lyrae Stars in the Galactic Bulge. Acta Astron. **61**, 1-23 (2011).

5. Christy, R.F. A Study of Pulsation in RR Lyrae Models. Astron. J. **69**, 536-537 (1964).

6. Smith, H.A. RR Lyrae Stars. Cambridge University Press, book (2004).

7. Bono, G., Caputo, F., Castellani, V., Marconi, M., Storm, J. & Degl'Innocenti, S. A pulsational approach to near-infrared and visual magnitudes of RR Lyr stars. Mon. Not. R. Astron. Soc. **344**, 1097-1106 (2003).

8. Stępień, K., Evolution of Cool Close Binaries – Approach to Contact, Acta Astron. **61**, 139-159 (2011).

9. Stępień, K., Evolutionary Status of Late-Type Contact Binaries, Acta Astron. **56**, 199-218 (2006).

10. Kunder, A. et al. Period Change Similarities Among the RR Lyrae Variables in Oosterhoff I and Oosterhoff II Globular Systems. Astron. J. **141**, 15-28 (2011).

11. Wilson, R.E. & Devinney, E.J. Realization of accurate close-binary light curves: application to MR Cygni. Astrophys. J. **166**, 605-620 (1971).

12. Van Hamme, W. & Wilson R.E. Third-Body parameters from whole light and velocity curves. Astrophys. J. **661**, 1129-1151 (2007).

13. Tognelli, E., Prada Moroni, P.G. & Deg'Innocenti, S. The Pisa pre-main sequence tracks and isochrones. A database covering a wide range of Z, Y, mass, and age values. Astron. Astrophys. **533**, A109 (2011).

14. Asplund, M., Grevesse, N., Sauval, A.J., Scott, P. The Chemical Composition of the Sun. Annu. Rev. of Astron. Astrophys. **47**, 481-522 (2009).

15. Prada Moroni, P.G. & Staniero, O. Very low-mass white dwarfs with a C-O core. Astron. Astrophys. **507**, 1575-1583 (2009).

16. Bono, G., Caputo, F., Cassisi, S., Incerpi, R. & Marconi, M. Metal-rich RR Lyrae variables. II. The pulsational scenario. Astrophys. J. **483**, 811-825 (1997).

17. Eggleton, P.P., Kiseleva-Eggleton, L., A Mechanism for Producing Short-Period Binaries, Astroph. Space Sci., **304**, 75-79 (2006)

18. Fabrycky, D., Tremaine, S., "Shrinking Binary and Planetary Orbits by Kozai Cycles with Tidal Friction", Astrophys. J., **669**, 1298-1315 (2007)

19. Tokovinin, A., Thomas, S., Sterzik, M., Udry, S., Tertiary companions to close spectroscopic binaries, Astron. Astrophys., **450**, 681-693, (2006)

**Acknowledgements** We gratefully acknowledge financial support for this work from the Chilean Center for Astrophysics FONDAP, the BASAL Centro de Astrofisica y Tecnologias Afines (CATA), NSF, Polish Ministry of Science (Ideas Plus), the Foundation for Polish Science (FOCUS, TEAM), and the GEMINI-CONICYT found. The OGLE project has received funding from the European Research Council "Advanced Grant" Program. It is a pleasure to thank the staff astronomers at Las Campanas and ESO Paranal who provided expert support in the data acquisition.

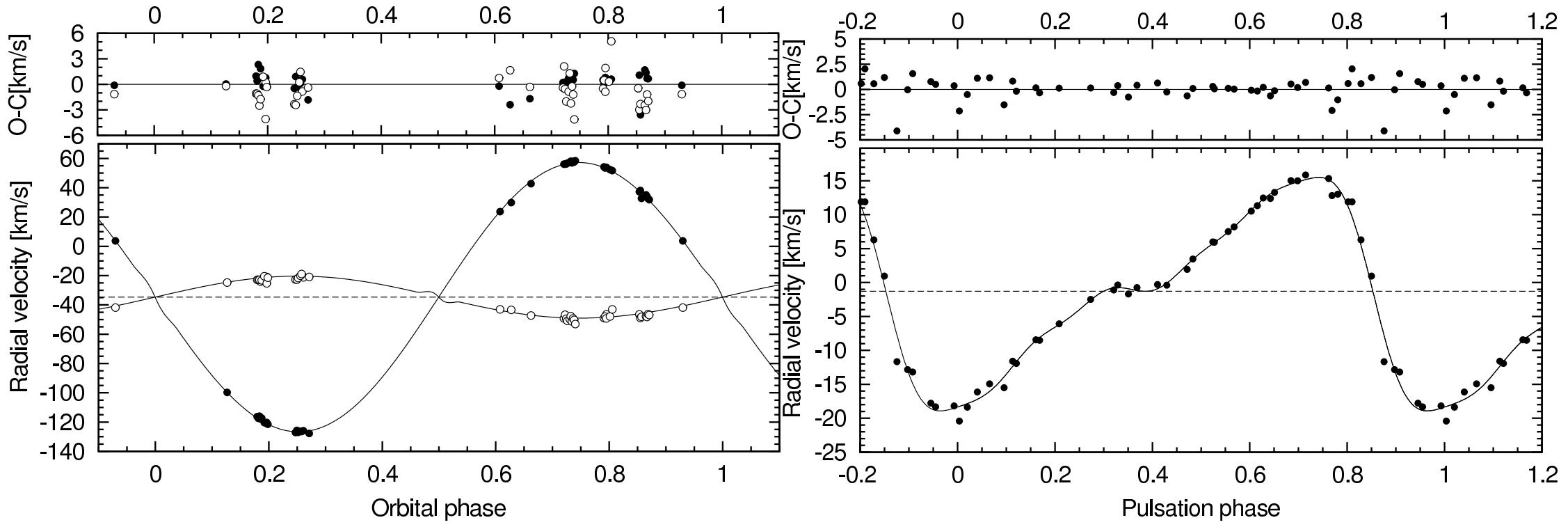

**Figure 1: Orbital motion of the two binary components, and the pulsational motion of the pulsating component of the OGLE-BLG-RRLYR-02792 system.** All individual radial velocities were

determined by the cross-correlation method using appropriate template spectra and the MIKE, and UVES spectra, yielding in all cases velocity accuracies better than 300 m/s (error bars smaller than the circles in the figure). Then, the orbit (mass ratio, systemic velocity, velocity amplitudes, eccentricity, and periastron passage) plus a Fourier series of order eight approximating the pulsation variations of the primary component, was fitted with a least squares method to the measured velocities. The resulting parameters are presented in Supplementary Table 1. The disentangled orbital radial velocity curves of both components of our binary system, and the pulsational radial velocity curve of the primary component are shown in the main a and b panels, respectively. Top, the residuals of the observed velocities (O) from the computed once (C).

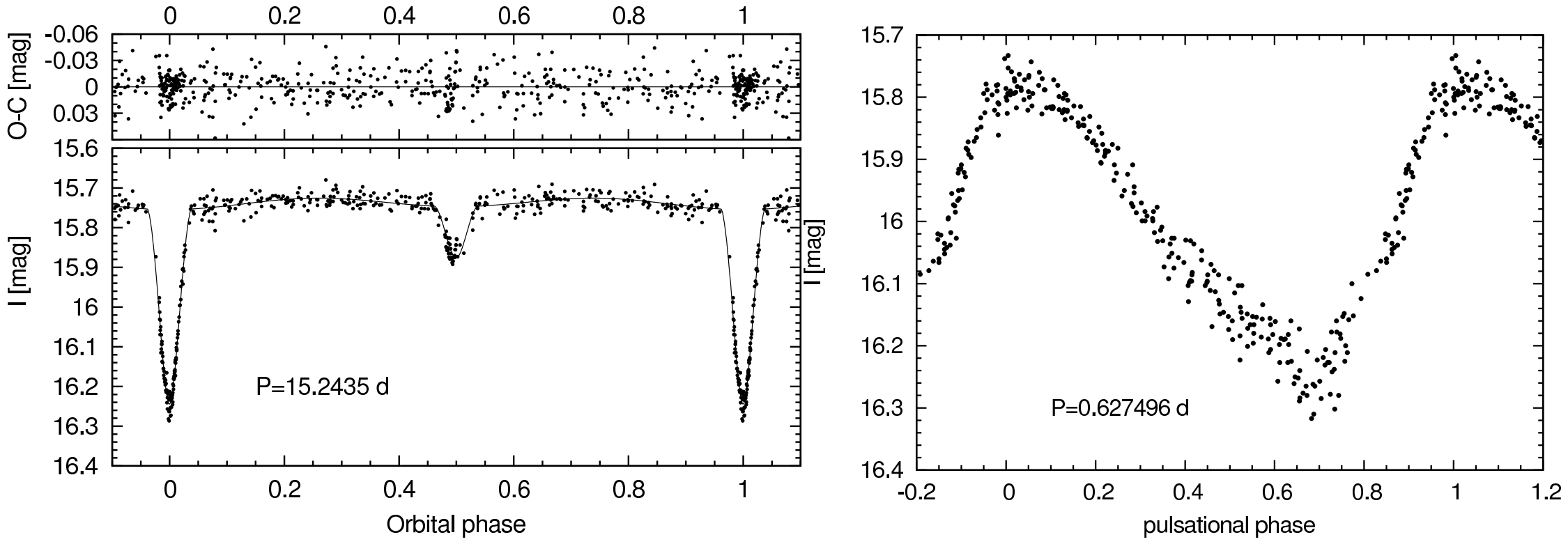

**Figure 2: Change of brightness of the binary system caused by the mutual eclipses, and the intrinsic change of the brightness of the primary component caused by its pulsations.**

The following final ephemeris for our system was derived from the OGLE photometric data:

$P = 15.24340 \pm 0.00021$ days  $T_0 = 2452108.3161 \pm 0.038$ days (orbital)

$P = 0.627496 \pm 0.000008$ days  $T_0 = 2455000.355 \pm 0.005$ days (pulsation)

Adopting the photometric ephemeris, and the mass ratio obtained from the analysis of the spectroscopic data (Figure 1) we model our spectroscopic and photometric observations using the 2007 version of the standard Wilson-Devinney (WD) code[11,12] . We accounted for the intrinsic photometric variations of the pulsating star in the system by fitting Fourier series of the order of 15 to the observations secured outside the eclipses and then subtract the corresponding variations in the eclipses in an iterative way, scaling the obtained fit according to the obtained WD model.

a: the orbital I-band light curve (617 epochs collected over 10 years) of the binary system OGLE-BLG-RRLYR-02792, after removal of the intrinsic brightness variation of the pulsating component together with the solution, as obtained with the Wilson Devinney code. The residuals of the observed magnitudes from the computed orbital light curve are also presented. b: the pulsational I-band light curve of the primary component of our binary system, folded on a pulsation period of 0.627496 days. The shape of the light curve is mimicking that of a classical RR Lyrae star.

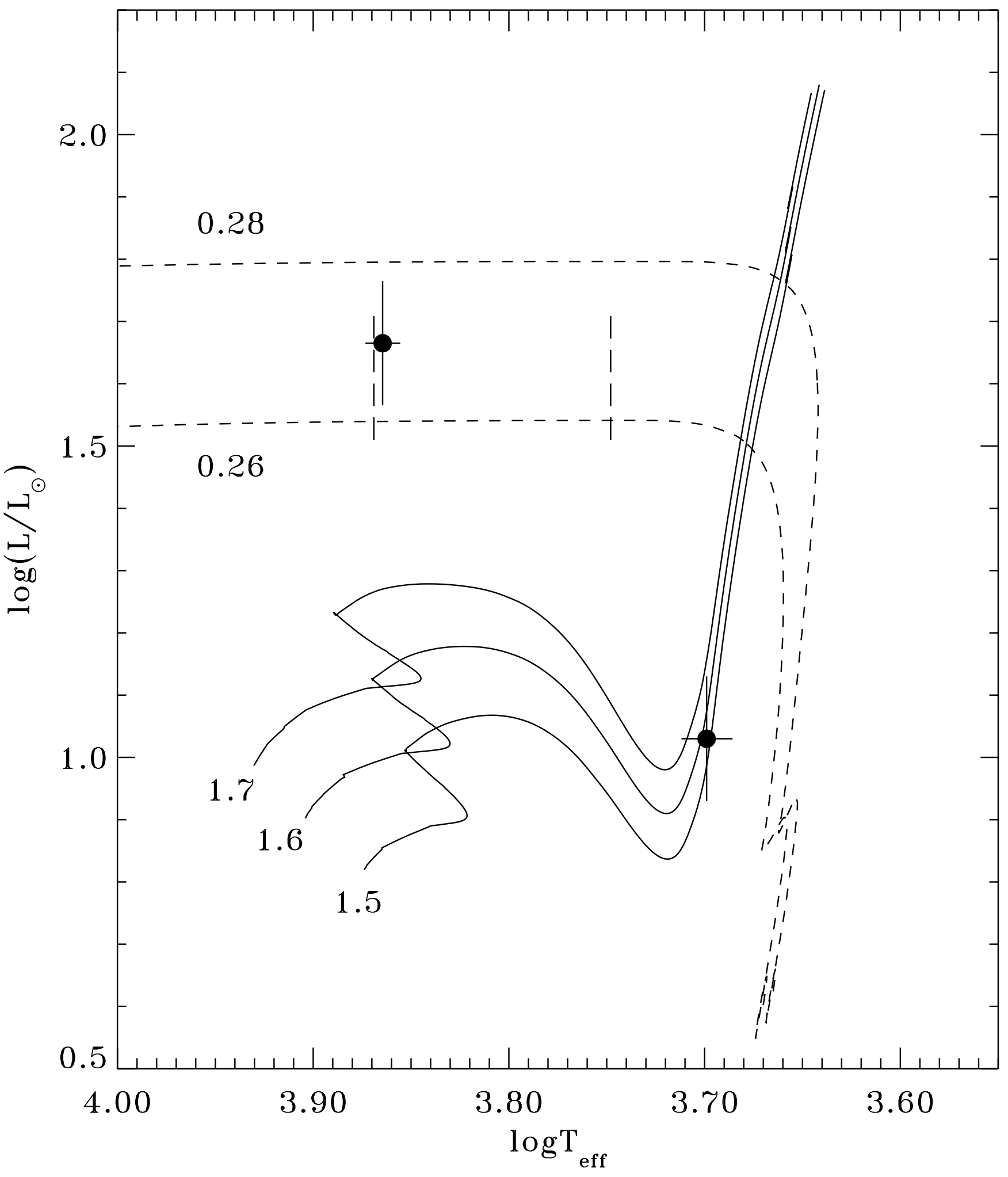

**Figure 3: Position of the two stars in the OGLE-BLG-RRLYR-02792 binary system on the H-R diagram.** The solid lines display evolutionary models computed with the most recent version of the FRANEC evolutionary code by adopting updated input physics[13]. Current evolutionary models were computed assuming a solar chemical composition (metals, Z=0.0129; helium, Y=0.274). We adopted the recent heavy-element solar mixture[14] and a mixing-length value of $\alpha$ = 1.74.

The dashed lines show the evolution of stellar structures with 0.26 and 0.28 $M_{\odot}$ computed following the standard evolution of a 1.4 $M_{\odot}$ stellar structure from the pre-main sequence up to the beginning of the RG phase. At $\log(L/R_{\odot})$=0.9 we applied an enhanced mass loss rate of about $10^{-7}$ $M_{\odot}$/yr until the final mass (i.e. 0.26 and 0.28 $M_{\odot}$) was approached. We computed the final evolutionary fate of these structures, at constant mass, down to the cooling phase of He-core WDs[15]. The two vertical dashed lines show the instability strip for typical RR Lyrae stars according to models for a solar chemical composition (Z=0.02, Y=0.28)[16], in which the pulsating component of our binary system is located. We adopted 300 K as the uncertainty of the calculated instability strip. Very good agreement between the evolutionary models and the observations is demonstrated.

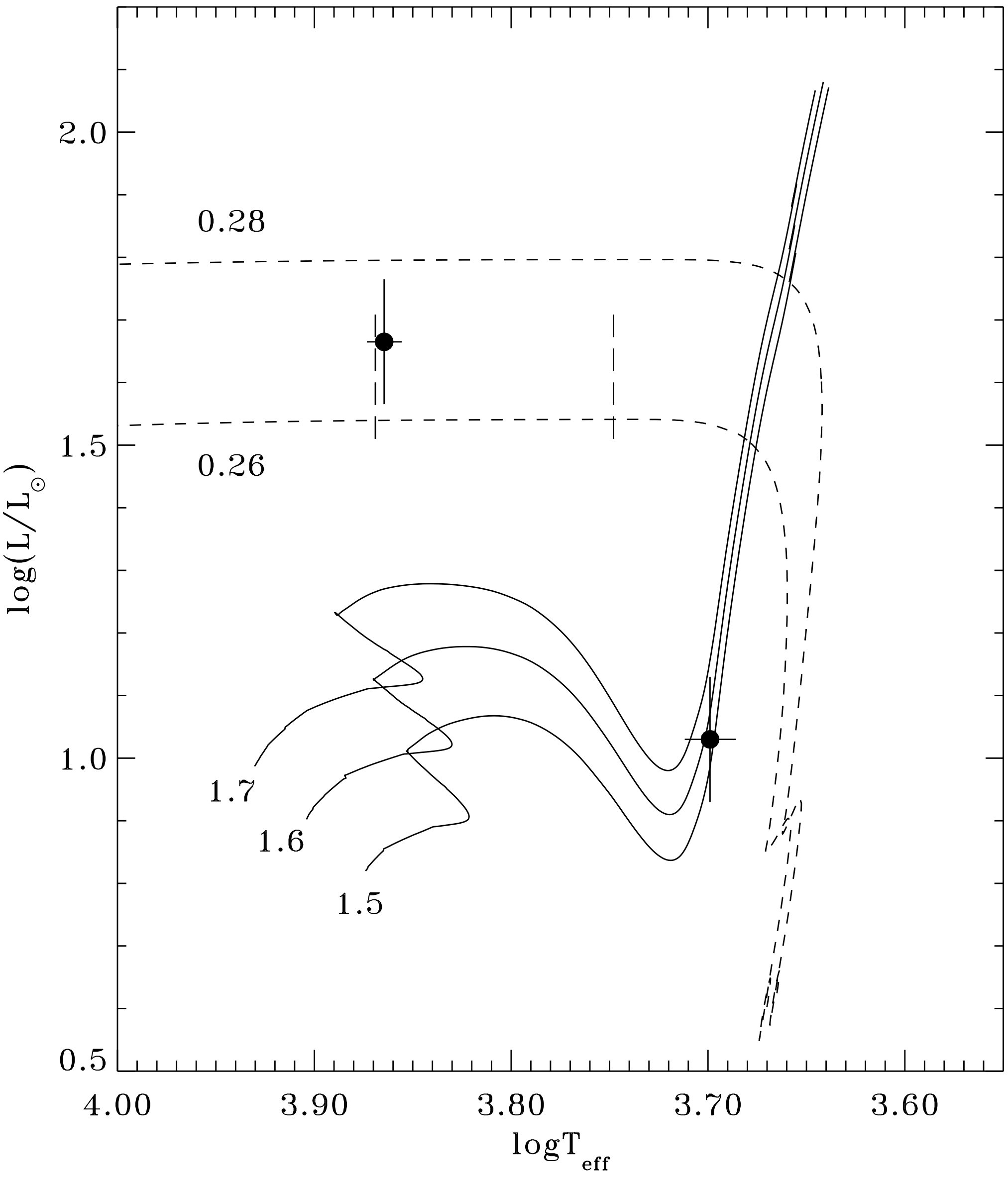

## Supplementary Information:

1) The essentials of the evolutionary model of close binaries.

We start from a detached binary with both components lying on ZAMS and possessing subphotospheric convection zones. Initial orbital periods are short enough to assure the synchronous rotation (currently, less than 3 d). During the first evolutionary phase, mass and angular momentum (AM) are lost by magnetized winds from both components. Any wind-wind interaction is neglected. The adopted mass loss rate (in sol.mass per year $M_{\odot\ 1,2} = -10^{-11}$ x $R^2_{\ 1,2}$ where radii are in solar units. This rate results from a fit to the observed mass loss rates extrapolated to rapid rotation of a star(when the magnetic activity reaches so called saturation level).

AM loss rate (in cgs units) is given $dH_{orb} / dt = -4.9 \times 10^{41} \times (R_1^{2}M_1 + R_2^{2}M_2) / P$, where radii are in solar units and period in days. This formula has been derived by KS and, again, applies to rapidly rotating stars.

The first phase takes typically several Gyr during which the components evolve accross MS. When the more massive component fills its inner critical Roche surface, the Roche lobe overflows (RLOF) occurs, followed by the rapid mass transfer to the companion. The process is assumed to be conservative, i.e. without any additional mass and AM loss. Mass transfer proceeds until both components reach thermal equilibrium. A short-period Algol system is formed. During the further evolution, mass and AM is still being lost by the winds but, at the same time, mass is slowly transferred from the presently low mass component (red giant/subgiant) to its companion. Mass transfer stops when the red component detaches from its Roche lobe because it is stripped of almost all hydrogen. The helium core with a thin hydrogen envelope moves to the region of the hot subdwarfs (crossing in the meantime the instability strip) whereas the other component leaves MS and moves towards the red giant branch. When it expands climbing up the branch, a common envelope develops.

2) Contamination of the RR Lyrae catalogs by binary evolution pulsators

In order to roughly estimate the contamination of RR Lyrae catalogs by binary evolution pulsating stars we assume that a half of stars are in eclipsing binary. In order to produce low mass stars mimicking RR Lyrae stars we need systems with the mass of the primary component in the range from 0.9 to 1.4 M_sun and having periods of 2-3 days. Taking into account that the Lidov-Kozai mechanism accompanied by the tidal friction can effectively shorten the period of about a half of systems with $3 < P < 10$, one can expect that about 1.5 % eclipsing systems can form binary evolution pulsating stars[9,17,18,19]. Assuming that about 20 % stars with masses between 0.8 and 0.9 M_sun will become RR Lyrae and using the initial mass function of Salpeter we estimate that there are 5 times more progenitors of RR Lyrae stars comparing to binary evolution pulsators. Taking into account that binary evolution pulsators cross the instability strip 100 times faster comparing to RR Lyrae we can expect the contamination ratio at the level of 0.2 %

Additional references :

17 - Eggleton, P.P., Kiseleva-Eggleton, L., A Mechanism for Producing Short-Period Binaries, Astroph. Space Sci., **304**, 75-79 (2006)

18 - Fabrycky, D., Tremaine, S., "Shrinking Binary and Planetary Orbits by Kozai Cycles with Tidal Friction", Astrophys. J., **669**, 1298-1315 (2007)

19. Tokovinin, A., Thomas, S., Sterzik, M., Udry, S., Tertiary companions to close spectroscopic binaries, Astron. Astrophys., **450**, 681-693, (2006)

Figure S1: H_alpha line observed in the spectrum of the primary component at the orbital phase of 0.25 (upper panel) and 0.75 (lower panel). Dashed and solid lines show the spectrum observed at the minimum and maximum pulsational velocity,respectively.

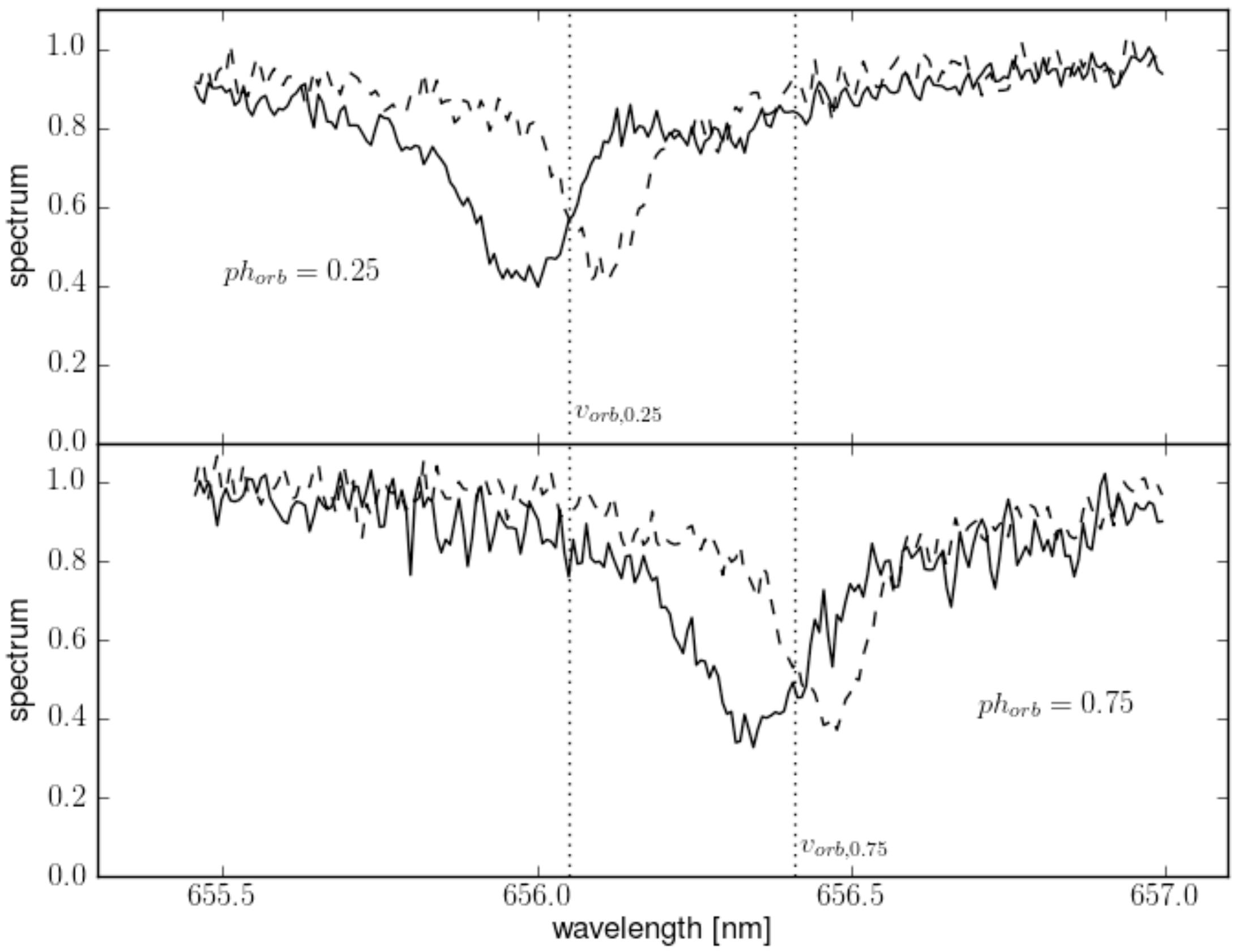

Figure S2: Location in the Mass – Radius plane of the two components of the OGLE-BLG-RRLYR-02792 binary system (filled circles together with corresponding 1 σ errors). The horizontal lines are the evolutionary models shown in Figure 3. The agreement between theory and observations is quite remarkable.

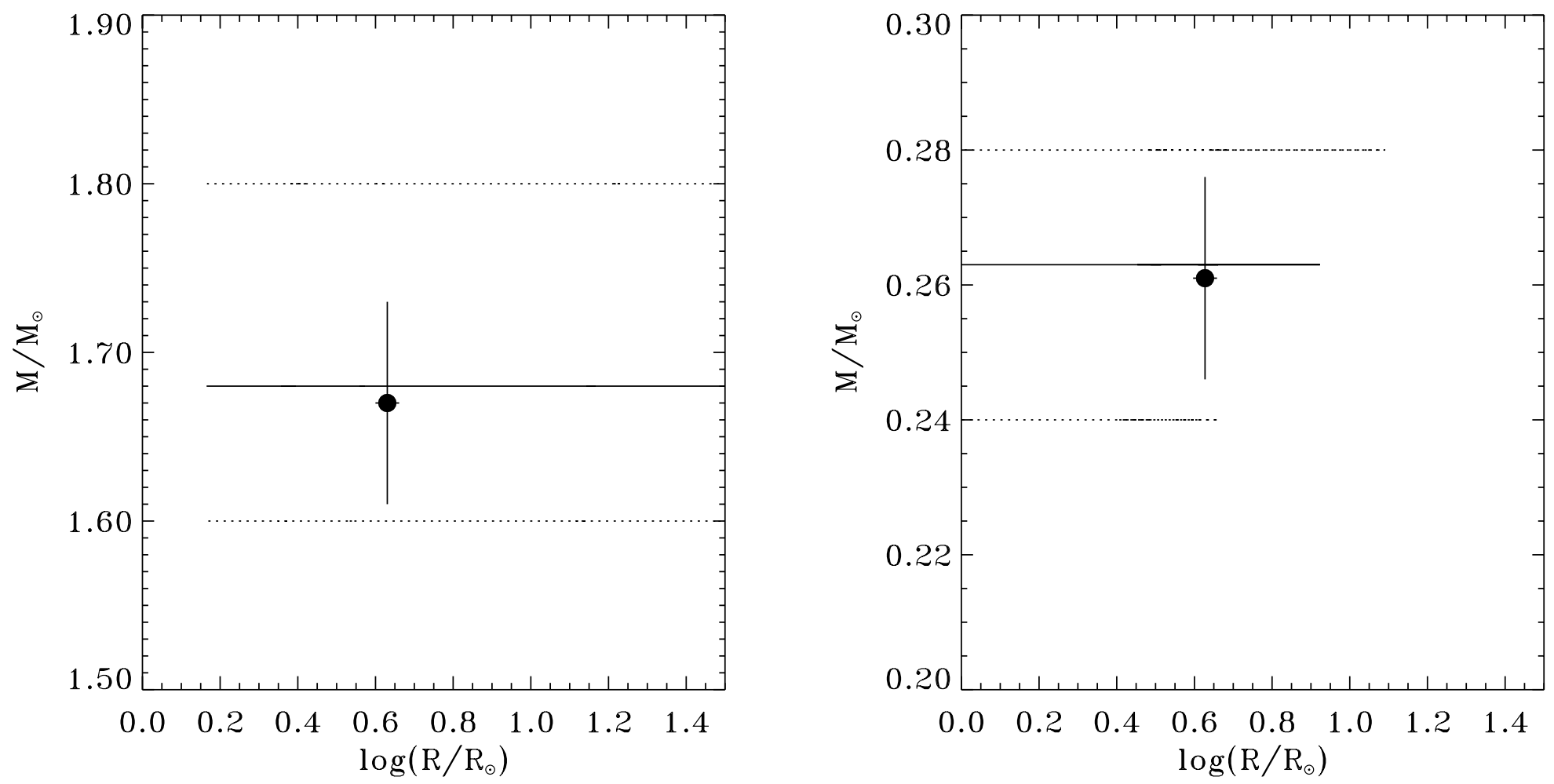

Supplementary Table 1. Orbital and physical parameters of the OGLE-BLG-RRLYR-02792 system (RA = $17^h47^m38^s.21$ DEC = $-35^o31'07''.1$, J2000.0) , together with their uncertainties as obtained from the modelling of the spectroscopic and photometric data.

A temperature of the red giant component of 5000 ± 150 K, (e.g. a typical temperature for a giant star with M = 1.67 $M_\odot$ and R = 4.24 $R_\odot$) was assumed. Then, the temperature of the pulsating component was calculated using the temperature ratio (for T2=5000 K, T1/T2 = 1.464 ± 0.010) obtained from the analysis of the light curve of our system. The mean I and V band magnitudes of the system measured outside the eclipses are 15.73 ± 0.01 mag and 17.08 ± 0.04 mag, respectively (L2/L1(I) = 0.298, L2/L1(V) = 0.161) . The measured pulsational amplitudes are $A_I$ = 0.36 ± 0.01 mag and $A_V$ =0.65 ± 0.05 mag. A finding chart for the system can be found on the OGLE Project webpage (http://ogle.astrouw.edu.pl)[4].

| Astrophysical parameters of the OGLE-BLG-RRLYR-02792 system | | |
|---|---|---|
| **Parameter** | **Primary (pulsating)** | **Secondary** |
| $M/M_{\odot}$ = mass | 0.261 ± 0.015 | 1.67 ± 0.06 |
| $R/R_{\odot}$ = *radius* | 4.24 ± 0.24 | 4.27 ± 0.31 |
| *K = velocity amplitude* | 91.83 ± 0.26 km/s | 14.31 ± 0.39 km/s |
| *T =effective temperature* | 7320 ± 160 K | 5000 ± 150 K (assumed) |
| e = eccentricity | 0.0072 ± 0.0029 | |
| ω=periastron passage | 277 ± 16 deg | |
| γ = systemic velocity | -34.7 ± 0.2 km/s | |
| $P_{ORB}$ $P_{PUL}$ = *periods* | 15.24340 ± 0.00021 days | 0.627496 ± 0.000008 days |
| $dP_{PUL}/dt$ = *period change* | (- 8.4 ± 2.6) x $10^{-6}$ days/year | |
| i = inclination | 83.4 ± 0.3 deg | |
| $a/R_{\odot}$ = orbit size | 32.20 ± 0.32 | |
| *q = mass ratio* | 6.42 ± 0.20 | |